\documentclass[prb,twocolumn,floats,aps]{revtex4}
\usepackage{graphicx,graphics,color,epsfig} 
\usepackage{bm}
\usepackage{amssymb}
\usepackage{amsmath}
\usepackage{amsfonts}

\newcommand{\mi}{{\,\mathrm{i}}}

\def\frac#1#2{{#1\over#2}}

\begin{document}

\title{Understanding disorder-induced zero-bias anomalies in systems with short-range interactions:  An atomic-limit perspective}
\author{R.\ Wortis}
\affiliation{Department of Physics \& Astronomy, Trent University,
1600 West Bank Dr., Peterborough ON, K9J 7B8, Canada}
\author{Lister Mulindwa}
\affiliation{University of Zambia,
Great East Road Campus,
P.O.\ Box 23279, Lusaka, Zambia}
\date{\today}

\begin{abstract}
Motivated by the novel electronic behaviors seen in transition metal oxides, we look for physical insight into disordered, strongly-correlated systems by exploring the atomic limit.  In recent work, the atomic limit has provided a useful reference point in systems with strong {\em local} interactions.  For comparison with experiments, the exploration of {\em non}local interactions is of interest.  In the atomic limit, both the case of on-site interactions alone and the case of infinite-range ($1/r$) interactions are well understood; however, not so the intervening possibilities.  Here we study the atomic limit of the extended Anderson-Hubbard model using classical Monte Carlo to calculate the single-particle density of states.  
We show that the combination of nearest-neighbor interactions and site disorder produce a zero-bias anomaly caused by residual charge ordering, and the addition of on-site interactions has a non-monotonic effect on the depth of this zero-bias anomaly.
A key conclusion is that the form of the density of states in this classical system strongly resembles density of states results obtained for the full extended Anderson-Hubbard model when $U<4V$.
\end{abstract}
\maketitle

\section{Introduction}
\label{sec-intro}

Transition metal oxides display a rich array of electronic behaviors many of which remain poorly understood.  
Because the highest occupied bands in these materials have strong $d$ and $f$ character, 
strong correlations are thought to play a significant role.  
Disorder is often present in these materials due to intrinsic and doped impurities as well as inhomogeneity arising from competing orders.
Understanding the influence of disorder in strongly-correlated systems is a significant challenge with implications for the characterization and control of transition metal oxides.

The single-particle density of states (DOS) is relevant to a wide range of materials properties and is a convenient point of contact between theory and experiment.
The combination of interactions and disorder is known to cause changes in the DOS near the Fermi level, and experiments are often compared with the two paradigms of the Efros-Shklovskii Coulomb gap\cite{Efros1975} and the Altshuler-Aronov zero-bias anomaly.\cite{Altshuler1985}
However, experimental data in strongly-correlated systems is frequently inconsistent with both of these pictures, and a better description of these systems is needed.\cite{Sarma1998,Richardella2010,Adhikary2012}

Significant progress has been made in understanding the behavior of the DOS near the Fermi level in systems with strong local interactions and strong site disorder.
Chiesa, et al\cite{Chiesa2008} demonstrated, using numerical techniques, the existence of a novel DOS suppression in the Anderson-Hubbard model.
Insight into the physical origin of this anomaly can be found by starting from the atomic limit.\cite{Wortis2010,Chen2010,Wortis2011, Chen2011}
The linear dependence of the width of the anomaly on the hopping amplitude, and independence from other parameters over a wide range of values, can be captured in a simple ensemble of two-site systems.\cite{Wortis2010}  As hopping between sites is turned on, the de-confinement of the electrons lowers their kinetic energy and results in a suppression of the DOS at the Fermi level.

More recently there has been numerical work\cite{Chen2012} exploring how the inclusion of {\it non}local interactions affects this picture of a kinetic-energy-driven zero-bias anomaly (ZBA).  
For weak nearest-neighbor repulsion, the kinetic-energy-driven ZBA persists in a renormalized form.
However, with stronger nonlocal interactions, in the parameter space where a clean extended Hubbard model displays charge density wave (CDW) order,\cite{Aichhorn2004} the behavior with disorder had a classical character.
While turning to the atomic limit to build understanding has proved useful in the Anderson-Hubbard model, the presence of nonlocal interactions presents new challenges.
When interactions are purely local, the atomic-limit DOS is simple to write down and at zero temperature there is no ZBA.  
In contrast, when nonlocal interactions are present even the clean atomic limit is nontrivial.\cite{Pawlowski2006,Pawlowski2008,Kapcia2011}

In this work we study the atomic limit of the extended Anderson-Hubbard model (EAHM) using classical Monte Carlo in order to address two groups of questions.
One topic is the Efros-Shklovskii Coulomb gap\cite{Efros1975} which arises in continuum insulating systems with $1/r$ interactions.
What happens to the Coulomb gap when the interactions are short ranged?  
Numerical work\cite{Efros1979JPC} has shown that the soft gap fills in when the interaction is exponentially screened, but the elegant analytic argument\cite{Efros1975} giving the energy dependence breaks down when interactions are cut off sharply at a finite distance.
Can we gain physical insight into the DOS suppression which remains?  
Moreover, the possibility of double occupancy has not been addressed.  
How is the Coulomb gap affected by onsite interactions?
A second topic is the full EAHM.  How similar is the behavior of the relatively simple atomic limit system to that with hopping?  Can the atomic limit case provide physical insight into the case with hopping?

We find that nearest-neighbor interactions alone result in a ZBA which arises from residual charge ordering and which, at strong disorder, has a width proportional to the strength of the nearest-neighbor interactions.  
The depth of the anomaly has an interesting non-monotonic dependence on the onsite interaction strength.
In addition, this simple classical system provides insight into many of the features found in the much harder quantum calculation on the full EAHM.\cite{Chen2012}

\section{Method}
\label{sec-meth}

Our extended Anderson-Hubbard model is a tight-binding Hamiltonian which includes the usual two Hubbard terms -- the hopping integral $t$ and onsite Coulomb repulsion $U$ -- plus two additional terms:  a nearest-neighbor Coulomb interaction  $V$ and disordered site potentials $\epsilon_i$.
\begin{eqnarray}
\mathcal{H}
&=& 
-t \sum_{\langle i,j \rangle,\sigma} {\hat c}_{i\sigma}^{\dag} {\hat c}_{j\sigma}
+ \sum_\mi U {\hat n}_{i\uparrow} {\hat n}_{i\downarrow} \nonumber \\
& & 
+ \sum_{\langle i,j \rangle} {V \over 2} {\hat n}_i {\hat n}_j 
+ \sum_{i,\sigma} \epsilon_i {\hat n}_{i\sigma},
\end{eqnarray}
${\hat c}_{i\sigma}^{\dag}$ is the creation operator for an electron with spin $\sigma$ at lattice site $i$,
${\hat n}_{i\sigma}={\hat c}_{i\sigma}^{\dag} {\hat c}_{i\sigma}$, and   
$\langle i,j \rangle$ refers to nearest neighbor pairs.
The site potentials $\epsilon_i$ are chosen from a flat distribution of width $\Delta$: $P(\epsilon_i)=\Theta(\Delta/2 - |\epsilon_i|)/\Delta$ where $\Theta$ is the Heaviside function.  

We focus on the atomic limit of this model, in which the hopping $t$ is set to zero.
In this limit, the physics is classical in the sense that the number of electrons on each site is always an integer.
In particular, the number of spin-up electrons on any given site can be either zero or one, and likewise for spin down.

We consider a two-dimensional square lattice with $N$ sites.  In particular the data presented are for a 20$\times$20 lattice with periodic boundary conditions.  
We restrict ourselves here to the case of half filling and equal spin populations.
We use a classical canonical-ensemble Monte Carlo simulation to calculate thermodynamic average properties, with a particular focus on the single-particle density of states (DOS).\cite{Mulindwa2011}

A simulation of a single disorder configuration begins by randomly assigning a potential $\epsilon_i$ to each site $i$ in the lattice according to the distribution above.  
The simulation proceeds by choosing an initial electronic configuration for the system and then proposing moves for electrons which are accepted with a probability consistent with detailed balance.  
After a period of equilibration, the DOS is calculated and averaged over Monte Carlo steps.  For any nonzero value of $\Delta$, multiple disorder configurations are simulated and the results are averaged.

To set the initial electronic configuration, $N/2$ spin-up electrons and $N/2$ spin-down electrons are placed.  
We generally do this randomly, but for comparison configurations such as Mott (1111) and checkerboard (2020) order were also considered.  

A single Monte Carlo step begins by randomly choosing an electron to propose moving.  Note that simply choosing a {\em site} at random creates a bias towards moving electrons on singly-occupied sites.  To address this, the program alternates between choosing electrons from singly and doubly occupied sites.  
Next, a possible new location for the electron is chosen randomly.  
If the energy difference for the move $\Delta E=E_{proposed}-E_{initial}$ is negative the move is always made, whereas if it is positive the move is made with probability $e^{-\Delta E/k_B T}$.

The time to reach equilibrium was determined by studying the time variation of the total energy and of the checkerboard order parameter $\Phi={1 \over 2}(n_A-n_B)$,\cite{Pawlowski2006} where $n_{A,B}$ are the numbers of electrons on the $A$ and $B$ sublattices of our bipartite lattice. 
Equilibration times increase rapidly as temperature is lowered, with weaker increases associated with lowering $U/V$ and $\Delta/V$.  For the results presented, equilibration times between 250,000 and 20 million sweeps were used, where a sweep is $N$ Monte Carlo steps.

After the equilibration period, the DOS is calculated once per sweep.  
In this classical system, there are only two energies at which a given site can contribute to the DOS.
An empty site contributes only at the energy to add an electron: $\omega=\epsilon_i+n_{nn}V-\mu$, the lower Hubbard orbital (LHO).
$n_{nn}$ is the number of electrons on neighboring sites; and $\mu={1 \over 2} U + 4V$ is the chemical potential at half filling.
Doubly occupied sites contribute only at $\omega=\epsilon_i+n_{nn}V+U-\mu$, the upper Hubbard orbital (UHO),
and singly occupied sites contribute with half weight each at both the LHU and UHO frequencies.
The total DOS is simply the sum of the contributions from all sites.

To calculate the thermal average DOS, we take an average over many Monte Carlo sweeps.  The energy and order-parameter autocorrelation time for different parameter sets were examined.  The autocorrelation time is longer at lower temperatures and lower values of $U/V$ and $\Delta/V$.  Data presented are averaged over 750,000 to 10 million sweeps, always including at least 100 autocorrelation times.
When $\Delta\ne 0$, a final average is taken over 1000 disorder configurations.

Consistency checks were done first by comparing with the Boltzmann distribution in small systems.  In addition, the variation of the specific heat with temperature at $\Delta=0$ was compared with earlier work in clean systems.\cite{Pawlowski2006}

\section{Results and Discussion}
\label{sec-rnd}

This section begins by considering the two limits of $V=0$ and $U=0$ to develop insight into the effects of $U$ and $V$ separately.  Next their combined effect is explored, and a comparison is made with work on the full EAHM.

\subsection{On-site interactions only}
\label{on-site-only}

\begin{figure}
\includegraphics[width=\columnwidth]{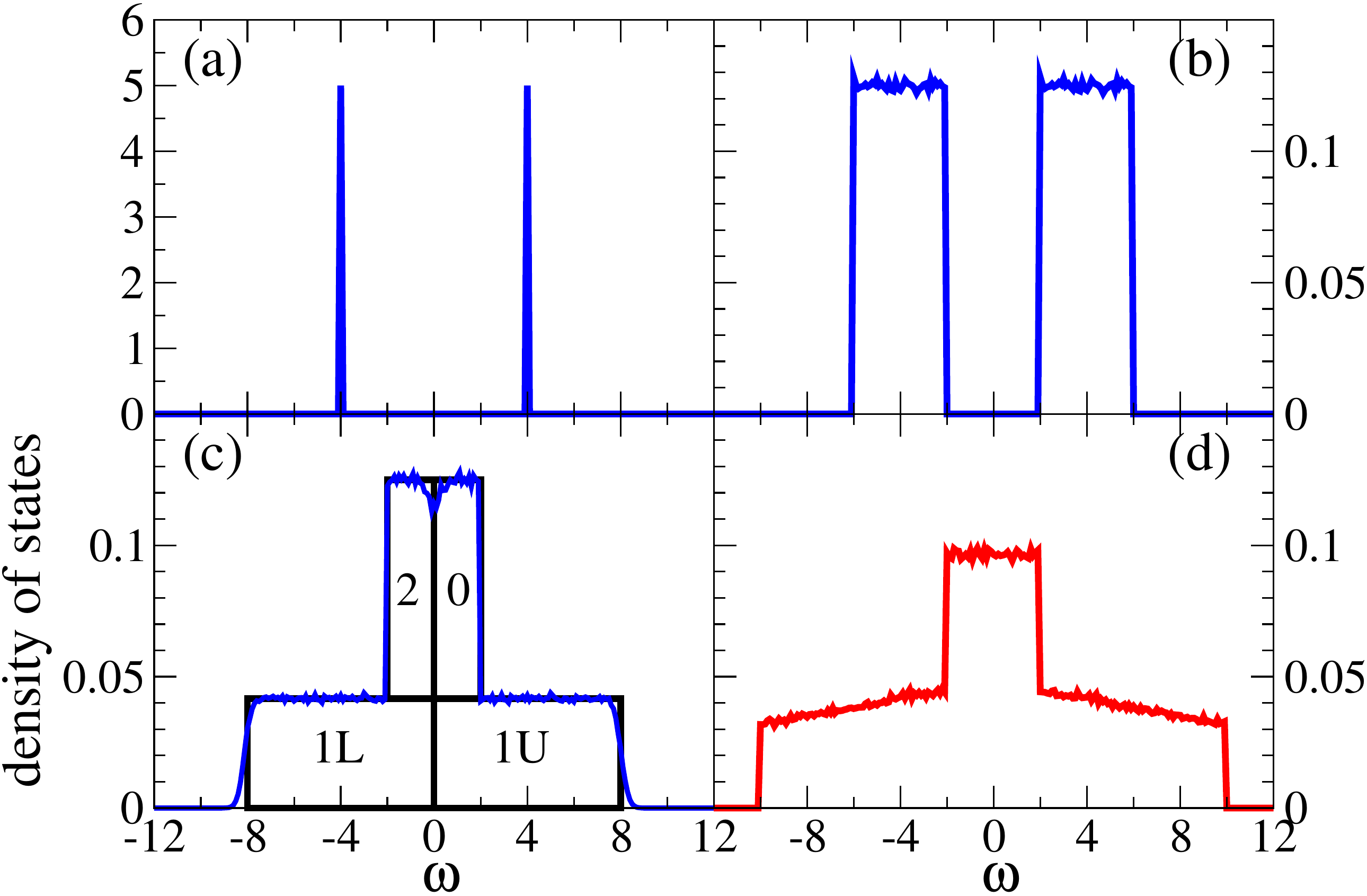}
\caption{\label{Uonly} Density of states for a purely local interaction: $U=8$ and $V=0$.  (a) $k_BT=0.1$ and $\Delta=0$, (b) $k_BT=0.1$ and $\Delta=4$, (c) $k_BT=0.1$ and $\Delta=12$, and (d) $k_BT=10$ and $\Delta=12$.}
\end{figure}

We begin our discussion by reviewing briefly the case with only local interactions.
In this case, the zero temperature DOS can be generated by simple arguments.  
In particular, the contribution of each site to the DOS is independent of the occupancies of any other sites.
When the site potential is greater than the chemical potential $\epsilon_i>\mu$, the site will be empty, and it will contribute to the DOS at $\omega=\epsilon_i-\mu$.
When $\epsilon_i< \mu-U$, the site will be doubly occupied, and it will contribute to the DOS at $\omega=\epsilon_i + U -\mu$.
The remaining sites will be singly occupied and will contribute to the DOS with half the weight at both energies.

In the clean system at half filling, all sites are singly occupied.
Therefore, the zero temperature DOS is two symmetric peaks at $-\mu$ and $U-\mu$ (Fig.\ \ref{Uonly} (a)).
When disorder is added, there is a continuous distribution of site potentials between $-\Delta/2$ and $+\Delta/2$.
So long as $\Delta<U$, each site remains singly occupied, but
the contribution to the DOS from each site falls at a slightly different frequency.
The two peaks of the clean case broaden into two bands (Fig.\ \ref{Uonly} (b)), each of width $\Delta$.
When $\Delta=U$, the two bands touch.
When $\Delta>U$, it might appear that the two bands overlap.  However, the key change is that some sites are now empty and some doubly occupied.  
These sites each make their DOS contribution at a single energy.  
Their contributions create a central plateau in the DOS which is three times the height of the shoulders.
In Fig.\ \ref{Uonly} (c), the labels indicate the occupancy of the sites contributing to the DOS in each energy range.

In Fig.\ \ref{Uonly} panels (a) and (b), the thermal average DOS obtained from our simulation closely matches the description just given.  In panel (c), however, there is a key difference: The DOS near $\omega=0$ is suppressed.  This is due to the fact that the Monte Carlo simulation is run, by necessity, at nonzero temperature.
At nonzero temperature, there is a temperature-driven ZBA which has been explored in detail.\cite{Wortis2011}
Briefly, at nonzero temperature a site with potential $\epsilon$ just above the chemical potential has a nonzero probability of being singly occupied.  
If it is singly occupied as opposed to empty, it no longer makes its full DOS contribution at $\epsilon-\mu$ but instead makes half its contribution at $\epsilon+U-\mu$.  
This therefore shifts spectral weight from the Fermi level to the top of the spectrum, where a small tail also appears.

As the temperature is increased, this thermal ZBA broadens and eventually the height of the central plateau declines.  When $k_B T$ is the dominant energy scale, each of the four occupancies--0, $\uparrow$, $\downarrow$ and 2--are equally likely at any site.  In this case, the DOS is simply the overlap of two flat bands of width $\Delta$ centered at $-\mu$ and at $U-\mu$.  Fig.\ \ref{Uonly} (d) shows the approach to this limit.

\subsection{Nearest-neighbor interactions only}
\label{nn-only}

\begin{figure}
\centering
\begin{tabular}{cc}
\parbox{0.6\columnwidth}{
\includegraphics[width=0.65\columnwidth]{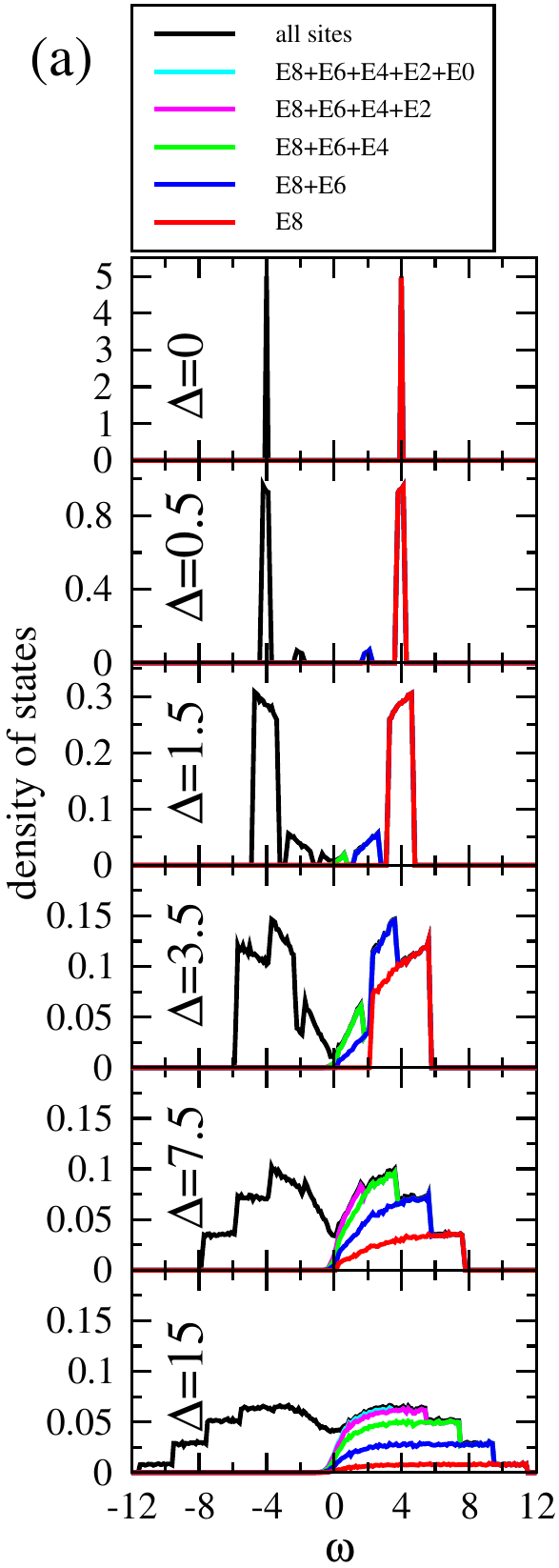} }
& \parbox{0.25\columnwidth}{\begin{tabular}{c}
\rule{0pt}{13 ex} \\
\rule{0pt}{14.2ex}
\includegraphics[width=0.75 in]{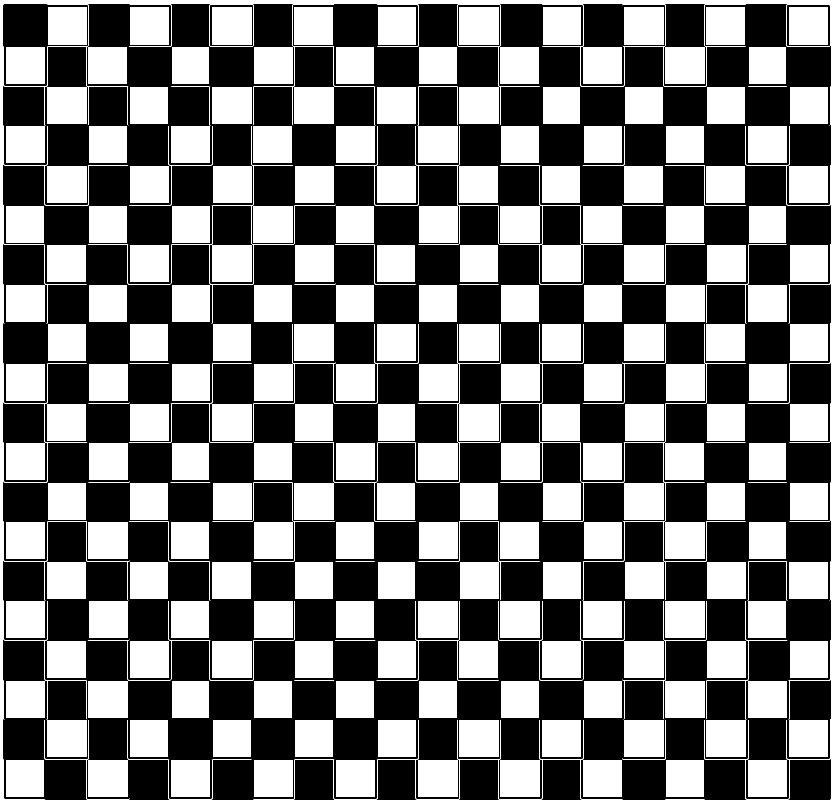} \\ 
\rule{0pt}{14.2ex}
\includegraphics[width=0.75 in]{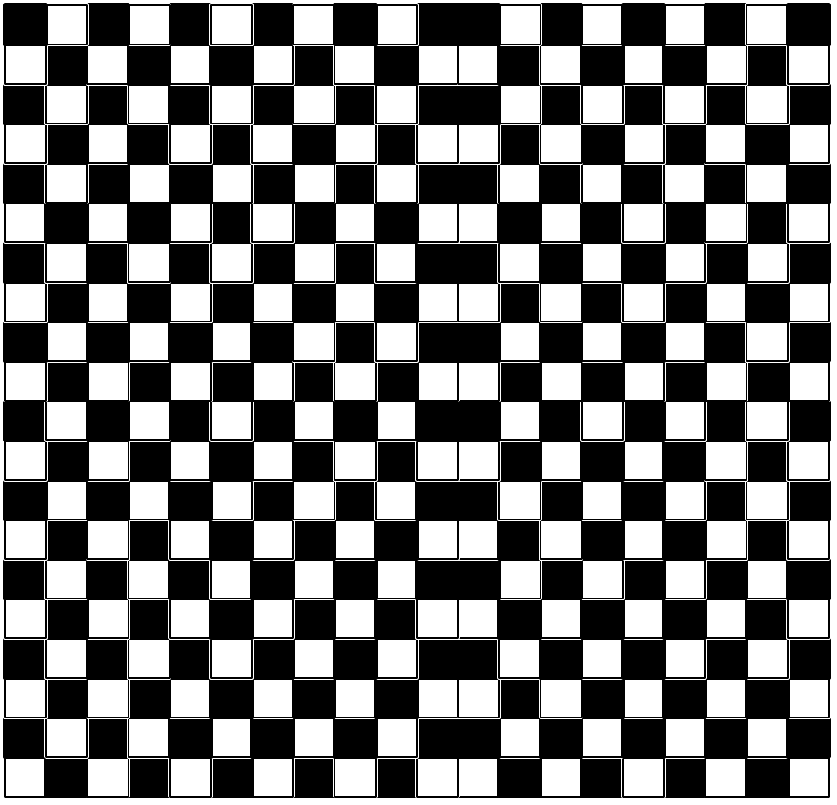} \\ 
\rule{0pt}{14.2ex}
\includegraphics[width=0.75 in]{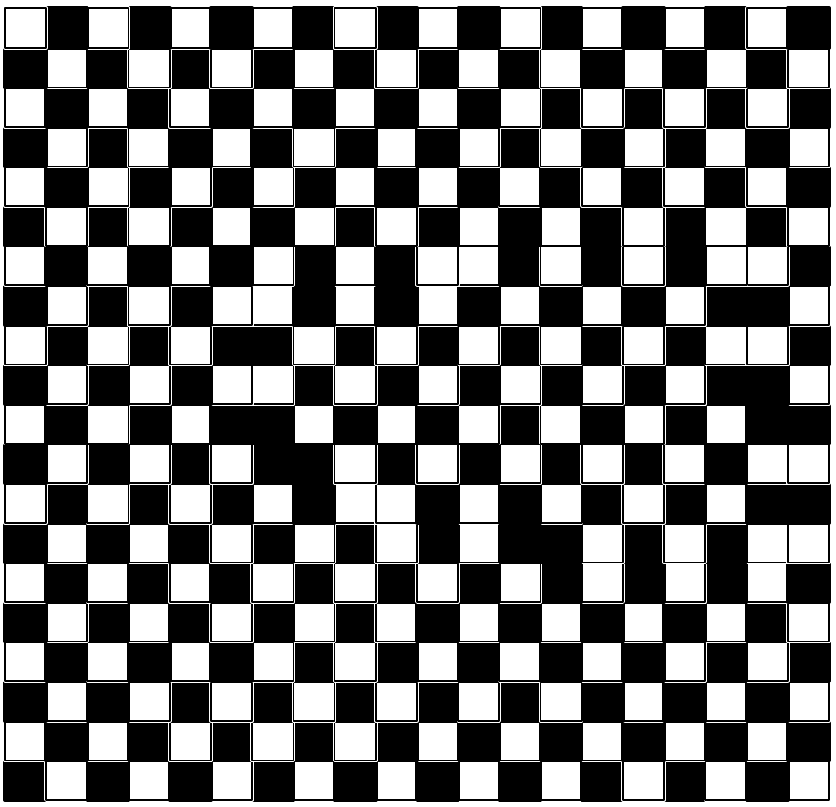} \\ 
\rule{0pt}{14.2ex}
\includegraphics[width=0.75 in]{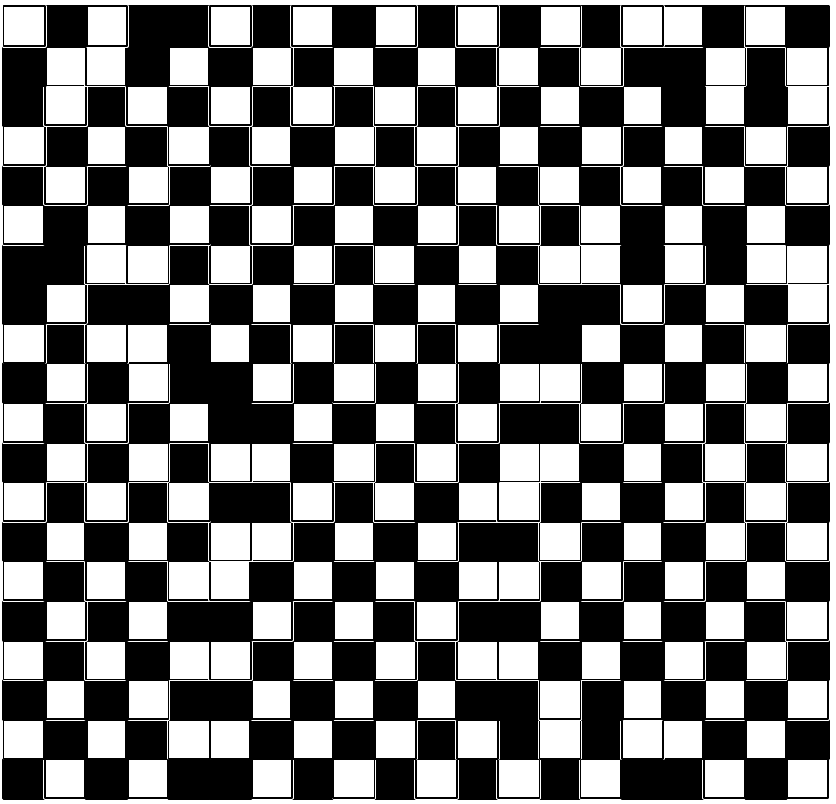} \\
\rule{0pt}{14.2ex}
\includegraphics[width=0.75 in]{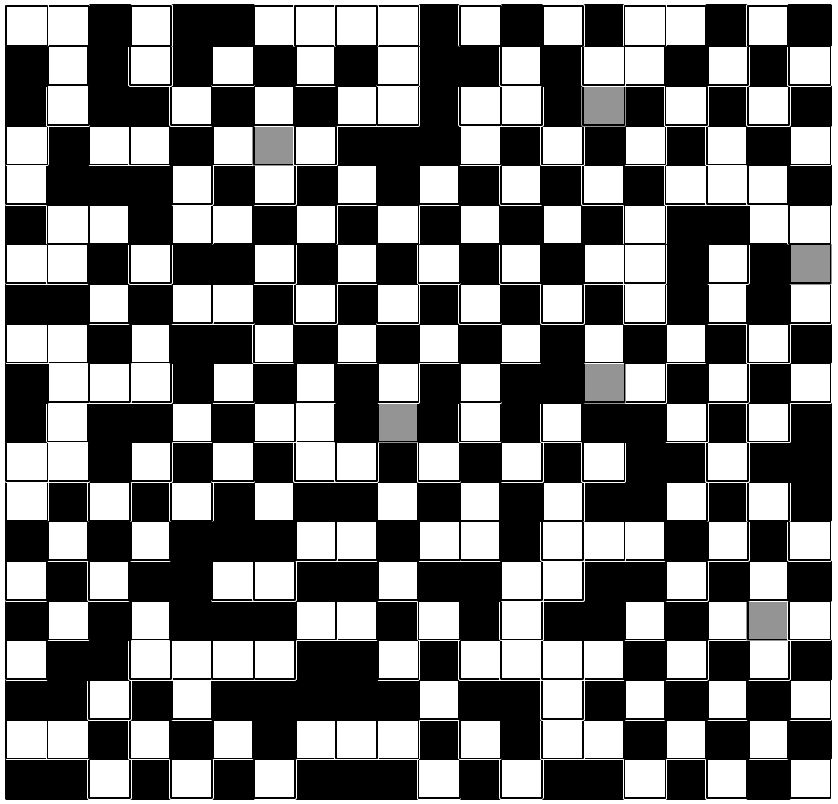} \\
\rule{0pt}{14.2ex}
\includegraphics[width=0.75 in]{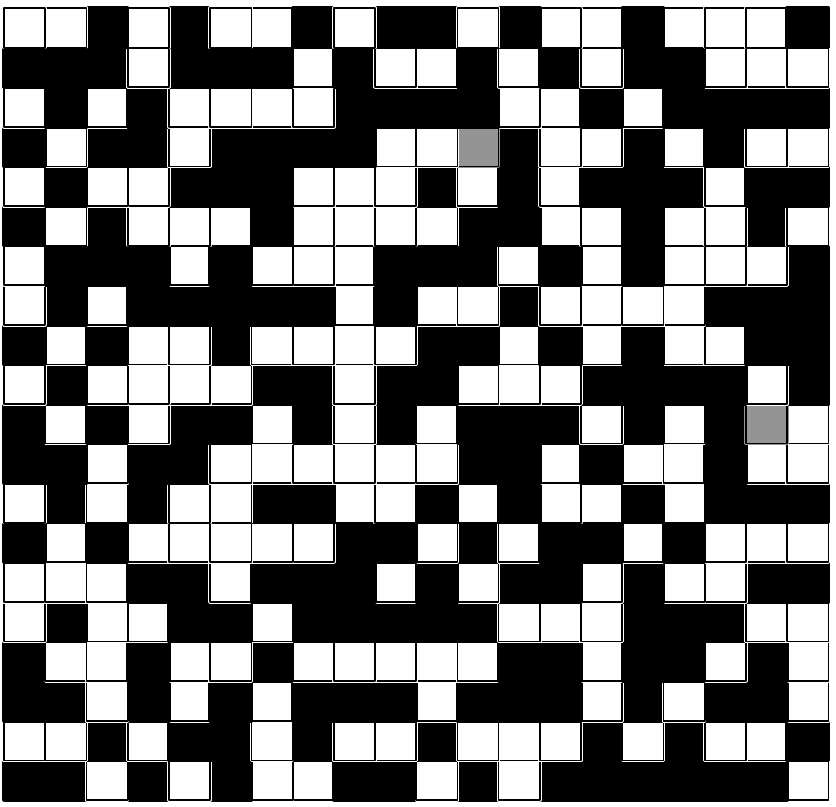} \\
\rule{0pt}{0ex}
\end{tabular}
}
\end{tabular}
\\
\includegraphics[width=\columnwidth]{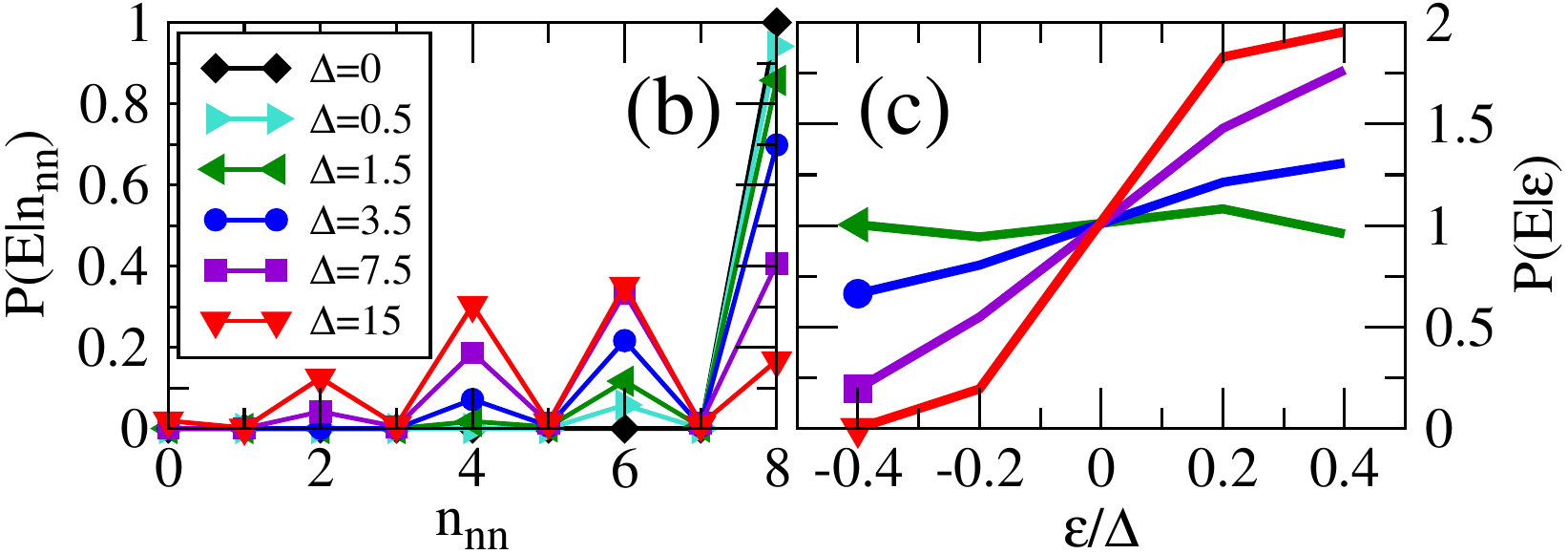} 
\caption{\label{Dfig} Variation of the density of states with increasing disorder strength.  (a) $U=0$, $V=1$, $k_BT=0.1$ and five disorder values as labeled.  To the right of each DOS plot is a typical configuration:  Black squares are empty, white squares are doubly occupied, and the small number of single occupied sites are grey.  (b) The probability that an empty site will have a given number of nearest neighbor electrons. (c) The probability that an empty site will have a given site potential.}
\end{figure}

\begin{figure}
\includegraphics[width=\columnwidth]{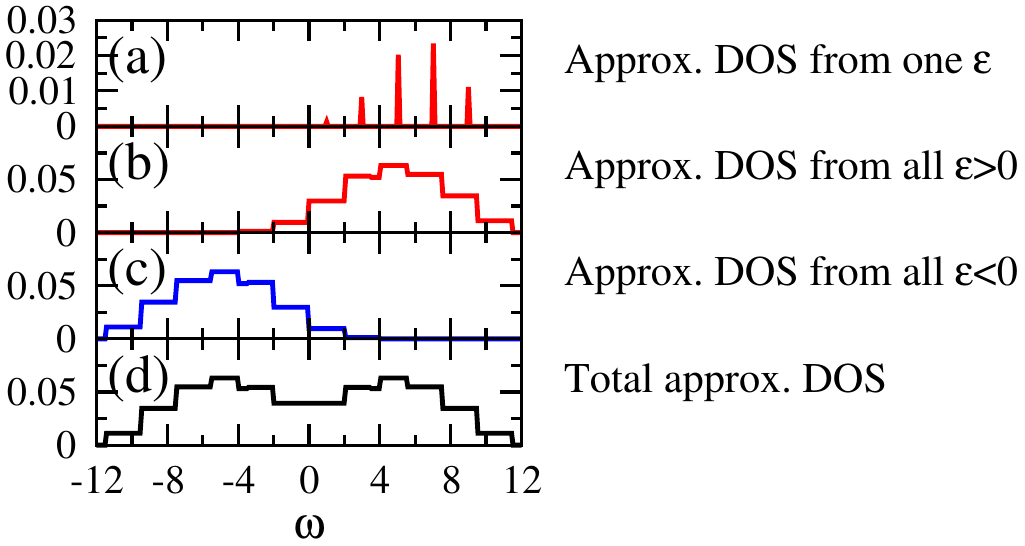}
\caption{\label{Dfig_extra} A simplified picture of the ZBA seen in the $\Delta=15$ panel of Fig.\ \ref{Dfig}(a).  (a) The DOS contribution of all sites with potential $\epsilon=5$ using $P({\rm E}|n_{nn})$ as shown in Fig.\ \ref{Dfig}(b).  (b) The DOS contribution from all sites with $\epsilon>0$ assuming that all these sites are empty and $P({\rm E}|n_{nn})$ is independent of $\epsilon$.  (c) The DOS contribution from all sites with $\epsilon<0$ assuming that all these sites are doubly occupied and $P({\rm D}|n_{nn})$ is independent of $\epsilon$.  (d) The total DOS within the above approximations.
}
\end{figure}

We now turn to the case of no on-site interactions.
This limit allows us to see the effect of the nearest-neighbor interactions most simply.

\subsubsection{Low temperature}

Fig.\ \ref{Dfig} (a) shows a sequence of DOS results with $U=0$, low temperatures $k_B T/V=0.1$ and increasing values of disorder $\Delta/V$.
The full DOS is indicated by the black line, and overlaid on this are colored lines showing the contributions from empty sites with specific numbers of nearest-neighbor electrons.  

At zero temperature, without on-site interactions, each site will be either empty (E) or doubly occupied (D).  
Nearest-neighbor interactions favor a checkerboard of E and D sites, as shown in the example configuration to the right of the $\Delta=0$ panel.
In a clean system at zero temperature, there are just two types of sites:  those which are empty and have eight electrons on neighboring sites (E8), and those which are doubly occupied with no electrons on neighboring sites (D0).  
The result is a DOS with just two peaks, separated by a charge-density-wave gap of $8V$.  
In the $\Delta=0$ panel, the red peak on the right is the contribution of empty sites with eight nearest-neighbor electrons (E8).

When a small amount of disorder is added ($\Delta=0.5$ and 1.5), the narrow peaks broaden into bands.
In addition, two other changes occur:  new peaks appear and the tops of the bands are not flat but slanted.  

New peaks appear because of the formation of domains in the checkerboard order.
While nearest-neighbor interactions continue to favor charge ordering, the disorder potential favors the placement of double occupancy at sites of below average potential.  
The competition between these results in domains.
The formation of domains for any nonzero disorder in two dimensions is a nontrivial point which has been examined in detail in the context of the random-field Ising model (RFIM),\cite{Natterman1998} onto which the atomic limit of the EAHM may be mapped when $U=0$.
For a domain of size $L$, the interaction energy cost is proportional to the length of the boundary, scaling as $L^{d-1}$ where $d$ is the dimension.
Meanwhile the potential energy savings of the domain scale as $L^{d/2}$.
When $L^{d/2} < L^{d-1}$, the system will remain ordered, 
whereas when $L^{d/2} > L^{d-1}$ domains will form.
A more detailed argument\cite{Aizenman1989} has shown that in two dimensions the RFIM forms domains for any nonzero value of disorder, with the size of the domains decreasing as the ratio of disorder to interaction strength increases.

Returning to our system, at the boundaries between domains, there are sites with neither zero nor eight nearest-neighbor electrons, resulting in the appearance of new bands in the DOS.
The example configuration to the right of the $\Delta/V=0.5$ panel shows a vertical domain boundary,
on which there are empty sites with only three neighbors doubly occupied (E6) and doubly-occupied sites with only three neighbors empty (D2). 
These sites contribute two new bands to the DOS.
Diagonal boundaries have a higher energy cost than vertical or horizontal boundaries and occur more frequently at higher disorder.  (See the $\Delta=1.5$ and $\Delta=3.5$ example configurations.)
Here there are E4 and D4 sites, resulting in another pair of new bands around $\omega=0$.

In addition to the broadening of bands and the appearance of new bands, disorder causes variation in height within each band.  

At low levels of disorder the bands slant almost linearly downward toward the Fermi level.
To understand this, consider the lowest band in the $\Delta=1.5$ panel.  
The low frequency side comes from the contribution of D sites at particularly low potential sites, whereas the high frequency side comes from the contribution of D sites at particularly high potential.  
The energetic preference for placing double occupancy at sites with below average potential is being accommodated by the formation of domains.  Therefore, double occupancy occurs more often at low potential sites than high potential sites, creating this slant in the DOS.
Likewise, high potential sites are more likely to be empty and hence the bands at positive frequencies are higher at the high frequency side.
As the disorder is further increased, the slanted bands overlap, resulting in a continuous DOS which has a minimum at zero frequency. 

At the two largest values of disorder shown, the bands are flat at the outside edge, retaining a downward slope only on the side closer to zero frequency.  
At high disorder strength, the idea of domains of checkerboard order becomes less useful in deciphering the pattern of occupation.  
In the example configuration at $\Delta/V=7.5$ the patches of checkerboard order are smaller, and by $\Delta=15$ many patches consist of just a single doubly-occupied (empty) site surrounded by four empty (doubly-occupied) sites.  
How can we tell whether these patches are occurring any more frequently than they would in a random arrangement?

A useful quantity to consider is the probability that an empty site will have a number $n_{nn}$ of nearest-neighbor electrons, $P({\rm E}|n_{nn})$.
For a clean system at zero temperature, this distribution is maximally asymmetric:  $P({\rm E}|n_{nn}=8)=1$ and $P({\rm E}|n_{nn}\ne 8)=0$.   (See $\Delta=0$ in Fig.\ \ref{Dfig}.)
On the other hand, at infinite temperature, where the occupancy of each site is entirely random, this distribution becomes symmetric about $n_{nn}=4$:  At infinite temperature, the probabilities are simply dictated by the number of ways of arranging the $n_{nn}$ neighboring electrons.  When only empty and doubly-occupied sites are considered, there are $2^4$ possible nearest-neighbor configurations, and
$P({\rm E}|8)=1/16$, $P({\rm E}|6)=4/16$, $P({\rm E}|4)=6/16$, etc.
The asymmetry of $P({\rm E}|n_{nn})$ is a convenient measure of the influence of nearest-neighbor interactions.

Fig.\ \ref{Dfig} (b) shows the evolution of $P({\rm E}|n_{nn})$ for the same disorder values shown in (a).  
There is a clear trend towards greater symmetry as the disorder is increased.  
Nonetheless, even at disorder strength $\Delta=15$, the distribution remains asymmetric.
Fig.\ \ref{Dfig_extra} provides a cartoon picture of how the asymmetry of $P({\rm E}|n_{nn})$ leads to a ZBA.
Panel (a) shows the contribution to the DOS from all sites with a particular site potential.  
This consists of five peaks, corresponding to the five possible numbers of nearest-neighbor electrons.
Panel (b) shows the contribution from all sites for which $\epsilon>0$ under the assumption that all these sites are empty and that $P({\rm E}|n_{nn})$ is independent of $\epsilon$.
The form is a series of steps.  Starting from the high frequency end, the step heights are proportional to $P({\rm E}|8)$, $P({\rm E}|8)+P({\rm E}|6)$, $P({\rm E}|8)+P({\rm E}|6)+P({\rm E}|4)$, etc.  
Panel (c) shows the contribution from sites with $\epsilon<0$, assuming in this case that they are all doubly occupied and that $P({\rm D}|n_{nn})$ is independent of $\epsilon$.  This is just the curve in (c) reflected across $\omega=0$.
Finally, panel (d) shows the sum of these two contributions.
Because $P({\rm E}|n_{nn})$ is skewed to the right, the contribution of the empty sites in panel (c) has an upward curvature around $\omega=0$.  Therefore, when it is added to the contribution from the doubly occupied sites, a dip occurs around $\omega=0$.  
If $P({\rm E}|n_{nn})$ and $P({\rm D}|n_{nn})$ were symmetric, the two contributions would sum to a uniform DOS.

The ZBA shown in the $\Delta/V=15$ panel of Fig.\ \ref{Dfig}(a) does not have sharp steps, but is instead smooth.  
In the discussion above we assumed that all sites with $\epsilon>0$ were empty.  This could be expressed as a statement that if a site is empty, the probability that it has a site potential $\epsilon$ is zero for $\epsilon<0$ and uniform for all $\epsilon>0$, i.e. $P({\rm E}|\epsilon)$ is a step function.
This is not, in fact, the case.  
Fig.\ \ref{Dfig}(c) shows the evolution of $P({\rm E}|\epsilon)$ for the same disorder values shown in (a).  
At low values of disorder, $P({\rm E}|\epsilon)$ is almost uniform across the full range of $\epsilon$.  
At intermediate values it develops a near-linear slant, but still extends over the full range of $\epsilon$.
(This is the origin of the slanted bands for $\Delta=0.5$, 1.5 and 3.5 in Fig.\ \ref{Dfig} (a).)
For strong disorder, $P({\rm E}|\epsilon)$ goes to zero for low values of $\epsilon$, flattens out again at high values of $\epsilon$, with a transition around $\epsilon=0$ which remains smooth and quite broad even at $\Delta=15$.
It is this gradual variation in $P({\rm E}|\epsilon)$ around $\epsilon=0$ which creates a smooth ZBA,
whereas sharp steps are still seen at the edges of the band because $P({\rm E}|\epsilon)$ becomes flat at large values of $\epsilon$.

To summarize, when $U=0$, at low disorder there is a hard charge-density-wave gap.
When disorder is strong enough that the bands in the DOS overlap, the DOS is sharply suppressed at the Fermi level due to the near-linear slant of $P({\rm E}|\epsilon)$, reflecting the preference for sites of relatively high potential to remain empty.
In this regime, the width of the local minimum around $\omega=0$ is related to the disorder strength $\Delta$, although complicated by the details of how the individual bands overlap.
Finally, when disorder is strong, a smoother, shallower ZBA comes from the persistent asymmetry of $P({\rm E}|n_{nn})$, reflecting the continued preference for empty sites to have larger numbers of nearest neighbors.
In this latter regime, the width of the ZBA is proportional to the interaction strength $V$.

With the two distributions $P({\rm E}|n_{nn})$ and $P({\rm E}|\epsilon)$ providing a framework, we can now proceed to explore the temperature dependence as well as the behavior when $U \ne 0$.

\subsubsection{Temperature dependence}

\begin{figure}
\includegraphics[width=\columnwidth]{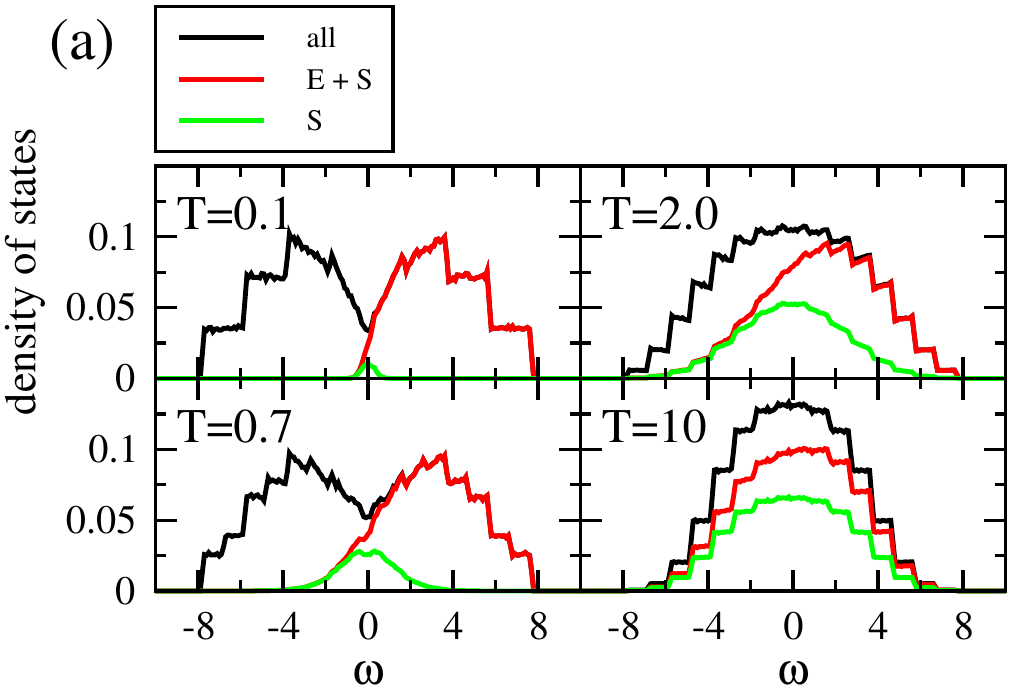} \\
\includegraphics[width=\columnwidth]{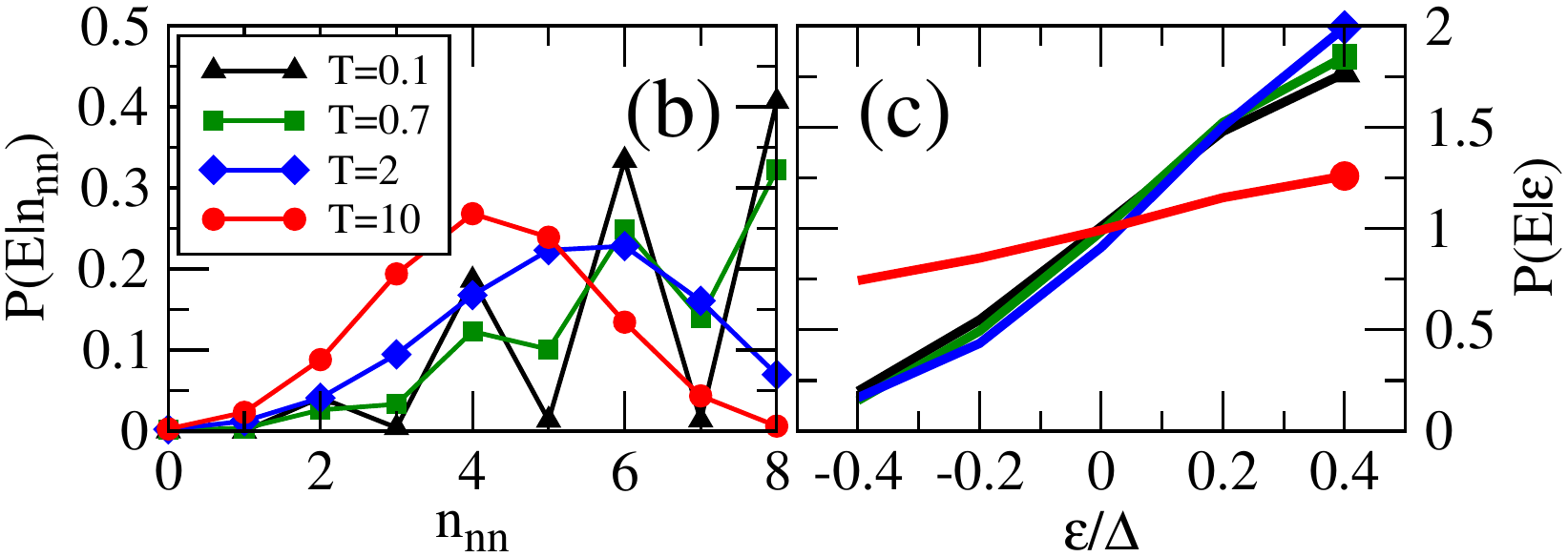} 
\caption{\label{Tfig} Variation of the density of states with increasing temperature.  (a) $\Delta=7.5$, $U=0$, $V=1$ and four temperatures as labeled.  (b) The probability that an empty site will have a given number of nearest neighbor electrons. (c) The probability that an empty site will have a given site potential.}
\end{figure}

Fig.\ \ref{Tfig}(a) shows the effect of increasing temperature, still for $U=0$.  
The first panel corresponds to the same parameters as the $\Delta=7.5$ panel of Fig.\ \ref{Dfig}(a).  
Here, the contribution to the DOS from the small number of singly occupied sites is shown in green.
The first major change introduced by increasing the temperature is an increase in the number of singly occupied (S) sites.  
This in turn causes two main changes in the DOS.
First, sites may now have odd numbers of nearest-neighbor electrons, increasing the number of possible values from five to nine.  
As a result there are more bands and hence more steps in the DOS.
Second, the ZBA comes from empty and doubly occupied sites, and since the fraction of empty and doubly occupied sites goes down the ZBA is reduced.

In addition to the changes associated with increasing single occupancy, increased temperature also dramatically reduces the asymmetry of both $P({\rm E}|n_{nn})$ and $P({\rm E}|\epsilon)$.  
Figs.\ \ref{Tfig}(b) and (c) show the evolution of $P({\rm E}|n_{nn})$ and $P({\rm E}|\epsilon)$ for the same temperatures as in (a).  
At high temperature, the probabilities for a site to have 0, $\uparrow$, $\downarrow$ or 2 electrons are all roughly equal, independent of the site potential or the local electron configuration.
$P({\rm E}|\epsilon)$ becomes flat because the occupancy of a site is no longer associated with the site's potential.
Moreover, the probability of a site having $n_{nn}$ nearest neighbors becomes proportional simply to the number of arrangements which give this value, resulting in a distribution which is peaked at $n_{nn}=4$. 
The reduced asymmetry in both $P({\rm E}|n_{nn})$ and $P({\rm E}|\epsilon)$ further washes out the ZBA.
Interestingly, at intermediate temperatures, $P({\rm E}|\epsilon)$ actually becomes slightly steeper,
probably due to a lowering of domain wall energies when singly-occupied sites are present.
This can be seen in the DOS in the increased slant of the steps at the edge of the band.

\subsection{Including both on-site and nearest-neighbor interactions}

Having considered the effects of $U$ and $V$ separately, we now address their combined effect.  
First, we will start at a fixed nonzero value of $U$ and gradually turn on $V$.
Then, we will start from a fixed nonzero value of $V$ and gradually turn on $U$.

\subsubsection{$V$ dependence}
\label{Vdep-disc}

\begin{figure}
\includegraphics[width=\columnwidth]{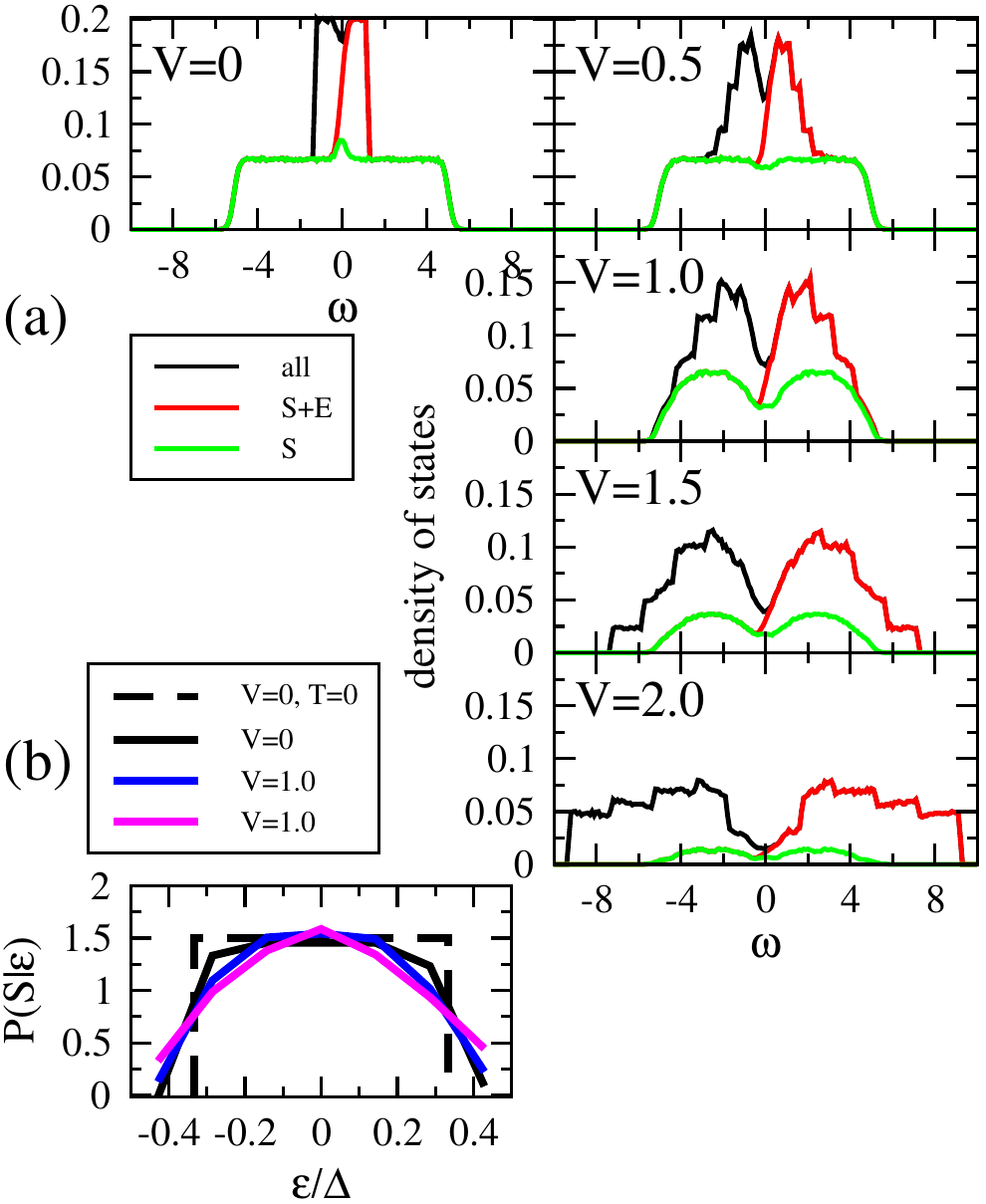}
\caption{\label{Vfig}  Variation of the density of states with increasing nearest-neighbor interaction strength.  (a) $\Delta=7.5$, $U=5$, $k_B T=0.1$ and five $V$ values as labeled.  (b) The probability that a singly occupied site will have a given site potential.}
\end{figure}

Fig.\ \ref{Vfig}(a) shows a sequence of DOS results starting from panel (c) of Fig.\ \ref{Uonly} and increasing nearest-neighbor interactions.
Fig.\ \ref{Vfig}(b) shows the evolution of $P(S|\epsilon)$ for the same parameter values as (a).
The $V=0$ case was discussed in Section \ref{on-site-only} A, but we revisit it here briefly in the context of the distribution $P(S|\epsilon)$.
At zero temperature and without nearest-neighbor interactions, $P(S|\epsilon)$ is uniform over the range $\mu - U < \epsilon<\mu$, as indicated by the dashed line in Fig.\ \ref{Vfig}(b).
As a result, the LHO and UHO $S$ contributions to the DOS run uniformly from $\omega=-U$ to $\omega=+U$, as shown in Fig.\ \ref{Uonly}(c).  
Nonzero temperature has the effect of spreading single occupancy over a broader range of site potentials, as seen in the $V=0$ curve in Fig.\ \ref{Vfig}(b).
The result is a slight increase in the $S$-site DOS contribution at $\omega=0$, due to the overlap of the tails in the distribution, but a net decrease in the DOS as discussed in \ref{on-site-only}.
When $V$ becomes nonzero, one effect is to generate a ZBA in the $E$ and $D$ contributions for all the reasons discussed in Section \ref{nn-only}.  
In addition, the $S$ sites also contribute to the ZBA.
Nearest-neighbor interactions narrow the range of site potentials at which single occupancy occurs:  For example, a site with $\epsilon =\mu-\delta$ which would have been singly occupied when $V=0$, if it has a large enough number $n_{nn}$ of nearest-neighbor electrons ($n_{nn}V-\delta>0$) will instead have a minimum grand potential when it is empty.
As $V$ increases, the distribution becomes rounded and slightly narrower, so the LHO and UHO $S$ contributions no longer sit flush with each other but pull back to form a ZBA.

Summarizing, when $U$ is nonzero and $V$ is turned on, the conversion of S sites to E and D sites results in a ZBA in the S contribution to the DOS which enhances the ZBA in the E and D contribution discussed above.
This S contribution to the ZBA results in an interesting re-entrant behavior when $U$ is varied.

\subsubsection{$U$ dependence}
\label{Udep-disc}

\begin{figure}
\includegraphics[width=\columnwidth]{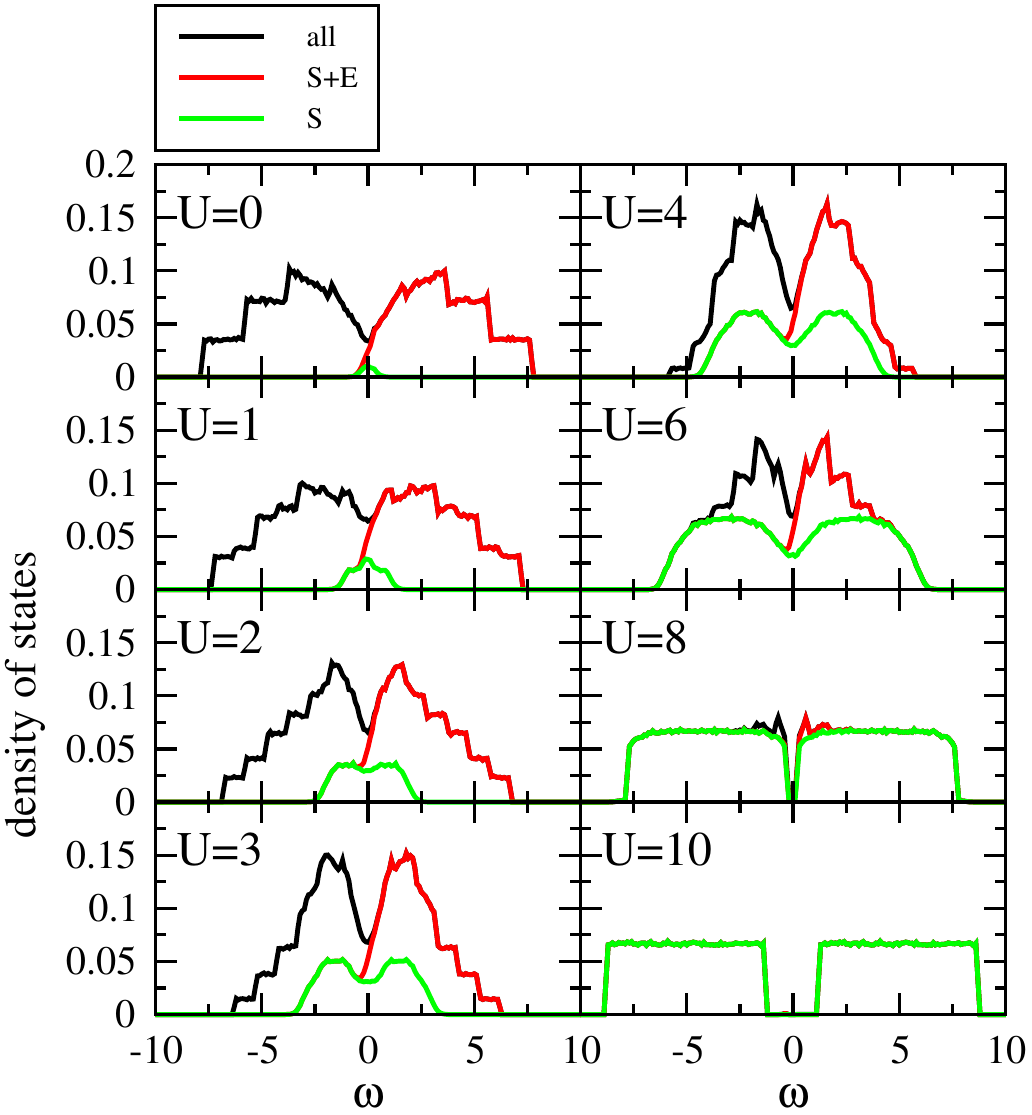}
\caption{\label{Ufig} Variation of the density of states with increasing on-site interaction strength.  $\Delta=7.5$, $V=1$, $k_BT=0.1$ and eight $U$ values as labeled.}
\end{figure}

Fig.\ \ref{Ufig} shows a sequence of DOS results starting from the $\Delta=7.5$ panel of Fig.\ \ref{Dfig}(a) and increasing onsite interactions.
We see similar features as in Fig.\ \ref{Vfig}(a) appearing here in the reverse order.  
When $U$ is zero, there is a ZBA in the E- and D-site contribution as discussed in Section \ref{nn-only}.  
The addition of on-site interactions generates single occupancy.
As in the case of nonzero temperature, this creates more steps at the edges of the band.
Meanwhile, at the center of the band, the ZBA is softened because the first sites to be singly occupied are near the Fermi level and hence contribute to the DOS near zero frequency.
Again, this contribution would be flat for $-U<\omega<+U$ if $k_BT=0$ and $V=0$.
At the lowest values of $U$ the thermal broadening effect dominates resulting in a slight peak in the $S$-site contribution at zero frequency, similar to that seen in Fig.\ \ref{Vfig}(a).
At higher $U$ values, the narrowing of $P(S|\epsilon)$ due to $V$ is resolved, resulting in a ZBA in the $S$-site contribution.
This causes the ZBA to become deeper again.  
Finally, at the largest values of $U$, single occupancy dominates and a Mott gap forms.  



\subsubsection{Comparison with quantum case}

Similarities between the atomic limit results presented here and the DOS of the EAHM when hopping is included\cite{Chen2012} can provide insight into the more computationally intensive quantum problem for some parameter ranges.  
In particular, Fig.\ 1 of Ref.\ [\onlinecite{Chen2012}] shows the evolution of the DOS with increasing nearest-neighbor interaction analogous to Fig.\ \ref{Vfig} here.  
The narrow $t$-dependent ZBA seen in Ref.\ [\onlinecite{Chen2012}] at low $V$ values is clearly different physics from what is seen here.
However, in both the quantum and the classical results a broad anomaly seen when $4V/U=1.2$ ($V=2.4t$ in Fig. 1 of Ref.\ [\onlinecite{Chen2012}] and $V=1.5$ in Fig.\ \ref{Vfig} of this paper).  
Moreover, at $4V/U=1.6$ the results with hopping show two broad slanted bands entirely consistent with the $E8$ and $D0$ bands in the atomic limit.
Why in the quantum case are only two bands seen, separated by a hard gap, while in our MC work the DOS contains multiple bands and remains continuous?   The quantum calculations used a very small sample size such that the domain size probably remained larger than the sample for all disorder configurations.

This insight helps resolve a point of confusion in Ref.\ [\onlinecite{Chen2012}].  In Fig.\ 5, of that work it was unclear whether the narrow ZBA which appeared at $U=4$ was the emergence of the kinetic-energy-driven ZBA for parameter values in which the clean system has CDW order or whether it was different physics.  
Comparing with our Fig.\ \ref{Ufig} it seems very likely that the narrow ZBAs seen in Ref.\ [\onlinecite{Chen2012}] Fig.\ 5 for $U=4$ and $U=6$ are a result of atomic-limit physics and distinct from the KE driven ZBA.

A quantum-classical crossover has also recently been noted in a Hartree-Fock study without local interactions.\cite{Amini2014}

\section{Conclusion}
\label{sec-concl}

With the broad goal of developing a theoretical framework for understanding zero-bias anomalies in disordered strongly-correlated systems, we have studied the relatively simple classical problem of the atomic limit of the extended Anderson-Hubbard model.  Our results shed light on two distinct issues.

First, regarding ZBAs in atomic limit systems, we have characterized the equivalent of the Efros-Shklovskii Couloumb gap in a tight-binding model with nearest-neighbor interactions.  
At low disorder strengths, the DOS suppression can be sharp with a width that is not set by $V$ but rather by details of the overlap of bands coming from groups of sites with the same number of nearest-neighbor electrons.  
At large disorder strengths, there is a smooth, shallow region of DOS suppression with an energy width proportional to the nearest-neighbor interaction strength $V$.  
The origin of this ZBA is residual charge ordering, namely the increased probability for empty sites to have more than the average number of nearest neighbors and for doubly occupied site to have fewer.  
While adding onsite interactions might be expected to suppress this ZBA by introducing single occupancy,
there is an interesting re-entrant behavior in which, after a weakening of the ZBA at low values of $U$, the ZBA deepens again at higher $U$ values.

Second, the DOS in this classical system bears a strong resemblance to results obtained from much more computationally intensive exact diagonalization studies of the full EAHM (i.e. with hopping) for $U<4V$.
The implication is that in this parameter range the interplay of interactions and disorder dominate the behavior while hopping has a negligible effect.  
We may conclude that the physical origin of the broad ZBA seen in the exact diagonalization studies in this regime is primarily the residual charge-ordering effect seen in our Monte Carlo studies.
In addition, our work provides a starting point for further exploration of the quantum case treating hopping as a small parameter.  
Along the lines of the ensemble of two-site systems used earlier to study the AHM,\cite{Wortis2010} an ensemble of small clusters could be constructed in which the frequency of specific occupancy patterns is set by those found in Monte Carlo studies of larger lattices.

While many subtleties arise when mapping tight-binding models onto real systems, this work is another step towards an alternative framework for interpreting tunneling and photoemission spectra on insulating strongly-correlated materials,
providing, for example, new diagnostics for determining the strengths of correlations in new materials.

\section*{Acknowledgments}
We acknowledge support by the National Science and Engineering Research Council (NSERC) of Canada.
This work was made possible by the facilities of the
Shared Hierarchical Academic Research Computing Network
(SHARCNET). 
R.W. thanks Michael Chomitz for helpful discussions.


\end{document}